\documentstyle[11pt,aaspp4]{article}

\begin{document}

\received{}
\accepted{}

\slugcomment{Submitted to The Astrophysical Journal}

\title{The Quasar Pair Q~1634+267 A, B and the Binary QSO vs. 
Dark Lens Hypotheses\footnotemark}

\footnotetext {Based on observations with the NASA/ESA {\it Hubble Space
Telescope}, obtained at the Space Telescope Science Institute, which is
operated by AURA, Inc., under NASA contract NAS 5-26555.}

\author{C. Y. Peng\altaffilmark{2}, C. D. Impey\altaffilmark{2}, 
	E. E. Falco\altaffilmark{3}, C. S. Kochanek\altaffilmark{3}, 
	J. Leh\'ar\altaffilmark{3}, B. A. McLeod\altaffilmark{3}, 
	H.-W. Rix\altaffilmark{2,4}, C.R. Keeton\altaffilmark{2},
	and J. A. Mu\~noz\altaffilmark{3}}

\altaffiltext{2}{Steward Observatory, The University of Arizona, Tucson, 
	AZ 85721}

\authoremail{cyp@as.arizona.edu}

\altaffiltext{3}{Harvard-Smithsonian Center for Astrophysics, 60 Garden St., 
	Cambridge, MA 02138}

\altaffiltext{4}{Current address: Max-Planck-Institut f\"{u}r Astronomie,
	Keonigstuhl 17, Heidelberg, D-69117, Germany}

\affil{email: cyp@as.arizona.edu, cimpey@as.arizona.edu, 
	falco@cfa.harvard.edu, ckochanek@cfa.harvard.edu, 
	jlehar@cfa.harvard.edu, bmcleod@cfa.harvard.edu, 
	rix@as.arizona.edu, ckeeton@as.arizona.edu, 
	jmunoz@cfa.harvard.edu}

\begin{abstract} 

Deep HST/NICMOS H (F160W) band observations of the $z=1.96$ quasar pair
Q~1634+267A,B reveal no signs of a lens galaxy to a 1$\sigma$ threshold of
$\simeq 22.5$ mag.  The minimum luminosity for a normal lens galaxy would be
a $6L^*$ galaxy at $z\simeq 0.5$, which is 650 times greater than our
detection threshold.  Our observation constrains the infrared mass-to-light
ratio of any putative, early-type, lens galaxy to $(M/L)_H
\gtrsim 690h_{65}$ ($1200h_{65}$) for $\Omega_0=0.1$ ($1.0$) and
$H_0=65h_{65}~\hbox{km~s}^{-1}~\hbox{Mpc}^{-1}$. We would expect to detect a
galaxy somewhere in the field because of the very strong Mg~{\sc ii}
absorption lines at $z=1.1262$ in the Q~1634+267 A spectrum, but the HST
H-band, I-band (F785LP) and V-band (F555W) images require that any
associated galaxy be very under-luminous $\lesssim 0.1L^*_H$ ($1.0L^*_I$)
if it lies within $\lesssim 40h_{65}^{-1}$ ($100h_{65}^{-1}$)~kpc from
Q~1634+267 A and B.

While the large image separation (3\farcs85) and the lack of a lens galaxy
strongly favor interpreting Q~1634+267A,B as a binary quasar system, the
spectral similarity remains a puzzle. We estimate that at most 0.06\% of
randomly selected quasar pairs would have spectra as similar to each other as
the spectra of Q~1634+267 A and B.  Moreover, spectral similarities observed
for the 14 known quasar pairs are significantly greater than would be
expected for an equivalent sample of randomly selected field quasars.
Depending on how strictly we define similarity, we estimate that only
0.01--3\% of randomly drawn samples of 14 quasar pairs would have as many
similar pairs as the observational sample.
\end{abstract}


\section{INTRODUCTION}

The number of strong gravitational lenses has grown tremendously in
the last five years, with well over 40 examples of strong
gravitational lenses produced by galaxies
(e.g.~\markcite{keeton1996}Keeton \& Kochanek 1996 and the CASTLES
webpage \footnote{http://cfa-www.harvard.edu/castles/}).  When the
separations are smaller than 3\farcs0, the lens galaxy has now been
detected in all but one system: the small-separation, high-luminosity
contrast lens Q~1208+1011.  The lens galaxies have normal photometric
properties and their mass-to-light ratios are typical of early-type
galaxies embedded in dark matter halos
(\markcite{kochanek1999b}Kochanek et al.\ 1999b,
\markcite{keeton1998}Keeton et al.\ 1998).  For the 18 systems with
separations larger than 3\farcs0 and nearly identical redshifts (see Table
1), a normal lens galaxy is detected in only four systems (RXJ~0911+0551,
Q~0957+561, HE~1104--1805 and MG~2016+112).  No lens galaxy has been
detected in any of the remaining 14 systems, with typical lower bounds on
the mass-to-light ratio of the lens about 10 times higher than those
measured for the normal lenses (\markcite{jackson1998}Jackson et al.\
1998).  Whether the {\it quasar pairs}\footnote{The definition of quasar
pairs also includes close physical systems which have been proven to be
non-lenses, but exclude known lenses.} with nearly identical redshifts
separated by 3\farcs0 to 10\farcs0 are examples of ``dark'' gravitational
lenses or binary quasars remains largely a mystery.  Proving either scenario
has very interesting ramifications on the dark mass distribution in the
universe, or the nature of interacting quasars (Schneider
1993\markcite{schneider1993}).  There are about 2 such quasar pairs for
every 1000 optically selected quasars (\markcite{hewett1998}Hewett et al.\
1998).

\markcite{kochanek1999a}Kochanek, Falco \& Mu\~noz (1999a) used a
comparison of the optical and radio properties of the quasar pairs to show
that the majority of the systems must be binary quasars.  First, five pairs
(PKS~1145--071, HS~1216+5032, Q~1343+2640, MGC~2214+3550 and
FIRST~J1643+315) are ``O$^2$R'' pairs in which only one of the quasars is a
radio source and we can be certain that the system is a binary quasar.
Second, even though most of the known lenses were radio selected and the
angular completeness of the radio lens surveys is better than that of the
optical lens surveys, there is no example of a ``dark'' radio lens (an
``O$^2$R$^2$'' pair).  These two facts are inconsistent with the dark lens
hypothesis.  Quantitatively, \markcite{kochanek1999a}Kochanek et al.\
(1999a) set a 2$\sigma$ (1$\sigma$) upper limit of 22\% (8\%) on the
fraction of the quasar pairs that could be gravitational lenses, the
most likely scenario being that none is a gravitational lens.

The high incidence and angular separations of the binary quasars can be
quantitatively explained in terms of merger induced quasar activity
(\markcite{kochanek1999a}Kochanek et al.\ 1999a).  The enhancement of the
quasar-quasar correlation function by a factor of $10^2$ on the angular
scales of the pairs is exactly as predicted by the locally observed
enhancement by a factor of $10$ in the quasar-galaxy correlation function on
the same physical scales (\markcite{fisher1996}Fisher et al.\ 1996,
\markcite{yee1987}Yee \& Green 1987, \markcite{french1983}French \& Gunn
1983).  The concentration of the pairs on scales smaller than 10\farcs0
($<50$~kpc) is a natural consequence of the need for a close passage and
tidal interactions to trigger renewed quasar activity.  The absence of
binaries on scales smaller than 3\farcs0 is a natural consequence of the
rapid increase in the orbital decay rate due to dynamical friction as the
binaries shrink.

These limits on the existence of dark lenses are, however, statistical.
While the maximum likelihood solution is that there are no dark lenses, the
statistical limit corresponds to allowing the existence of 3 dark lenses at
the 2$\sigma$ limit.  The existence of even 3 dark lenses would mean that
the densities of dark lens halos and normal clusters are the same on these
mass scales (\markcite{maoz1997}Maoz et al.\ 1997,
\markcite{kochanek1995}Kochanek et al.\ 1995,
\markcite{wambsganss1995}Wambsganss et al.\ 1995).  Moreover, the binary
hypothesis must somehow explain the remarkable similarity of the quasar pair
spectra in the optical (e.g. \markcite{hawkins1997}Hawkins et al.\ 1997,
\markcite{hewett1989}Hewett et al.\ 1989,
\markcite{steidel1991}Steidel \& Sargent 1991,
\markcite{michalitsianos1997}Michalitsianos et al.\ 1997). 

Q~1634+267A,B remains one of the best candidates for exploring the issue of
dark lenses and binary quasars.  After the discovery by
\markcite{sramek1978}Sramek \& Weedman (1978) in a slitless spectroscopy
survey, \markcite{djorgovski1984}Djorgovski \& Spinrad (1984) and
\markcite{turner1988}Turner et al.\ (1988) found the optical spectra of the
two quasars to be very similar, with an upper limit on the velocity
difference of $150~\hbox{km~s}^{-1}$, at the redshift of $z=1.961$.  The
flux ratio of the pair was 4.4 at R-band, although in images with seeing
comparable to the image separation (\markcite{djorgovski1984}Djorgovski \&
Spinrad 1984).  No lens galaxy has been detected down to limits of
$K\approx$ 21.5 (\markcite{steidel1991}Steidel \& Sargent 1991) and
$R=23.5$ (\markcite{djorgovski}Djorgovski \& Spinrad 1984).

Steidel \& Sargent (1991; hereafter SS91) used the Palomar 5-meter telescope
to obtain very high signal-to-noise spectra of Q~1634+267 that illustrate the
striking similarity of the spectra of components A and B, modulo a constant
scaling factor of 3.28.  They also found a number of absorption lines in the
brighter component A, most notably the Mg~{\sc ii} doublet
$\lambda\lambda$2796,2803 at $z=1.1262$, which perhaps hinted at the
presence of a lens galaxy.  When the emission line spectra of A and B were
compared in detail, there were slight differences in the Ly~$\alpha$ + {\sc
N~v}, Si~{\sc iv} + {\sc O~iv]}, and {\sc C~iv} line profiles, and the line
velocities were shifted by as much as 300--500~km~s$^{-1}$
(\markcite{small1997}Small et al.\ 1997).  The time delay created by the
different travel times for the two rays can lead to spectral differences
under the lens hypothesis, if the source quasar is variable.  The temporal
variations of quasar spectra have been examined by
\markcite{filippenko1989}Filippenko (1989), \markcite{wisotzki1995}Wisotzki
et al.\ (1995),
\markcite{impey1996}Impey et al.\ (1996), and
\markcite{small1997}Small et al.\ (1997).  In particular,
\markcite{small1997}Small et al.\ (1997) demonstrated that the
spectral differences observed in Q~1634+267A,B were typical of the
differences seen in spectra of the same quasar taken after an interval
of several years.

With the advent of a second generation of instruments on the Hubble
Space Telescope (HST), NICMOS has offered unique capabilities to
discover and identify missing lens candidates by observing them at
rest optical wavelengths where they are bright, while providing
unprecedented high angular resolution in the infrared necessary for
studying these systems in detail.  The
CfA/Arizona~Space~Telescope~Lens~Survey (CASTLES) project is currently
conducting a study of roughly 40 gravitational lens systems and
candidates using optical/near infrared imaging with the HST.  The
selection sample consists of ``simple'' systems where the strong
lensing is believed to be caused predominantly by a single galaxy.
Some of the goals are: to place tighter constraints on the lens
geometry; to study the mass and light distribution of the lens galaxy;
to study the evolutionary history and environment of the lens galaxy;
to better constrain cosmological parameters through lens modeling and
statistics, and finally, in some cases to discover the missing lens
galaxy.

In Section 2 we present our observations of Q~1634+267A,B, the photometry,
and the determination of the magnitude limit at the anticipated lens
position.  We also present new limits on the radio flux of Q~1634+267A,B and
two other quasar pairs. Section 3 presents parameters derived for a
hypothesized lens galaxy from a simple isothermal sphere model.  We discuss
the properties of the metal line absorber at $z=1.1262$ in Section 4, and
the statistical similarity of quasar spectra in Section 5.  Conclusions
follow in Section 6.  We adopt $H_0=65~\hbox{km~s}^{-1}~\hbox{Mpc}^{-1}$
throughout the paper and display key results for both $\Omega_0=0.1$ and
$\Omega_0=1$.

\section{OBSERVATIONS AND ANALYSIS}

\subsection{Observations}

We observed Q~1634+267 on 15 August 1997 with HST using NICMOS Camera 2 (NIC2)
and the $H$ (F160W) filter.  We obtained four 690-second integrations in a
10.5-pixel dither pattern, for a total integration time of 2560 seconds. We
reduced the data with our own software package ``nicred''
(\markcite{mcleod1997}McLeod 1997; \markcite{lehar1999} Leh\'ar et al.\
1999).  Figure 1 shows the reduced, combined image of the field of
Q~1634+267.  Q~1634+267 A and B are approximately centered in the NIC2
field, and a star is visible at the edge of the $\sim 19\arcsec\times
19\arcsec$ field of view.

\subsection{Astrometry and Photometry}

The plate scales of the NICMOS cameras changed as the IR array
underwent thermal expansion. The variation was monitored
\markcite{cox1998}(Cox et al.\ 1998), and the NIC2 plate scales were
0\farcs0760926 and 0\farcs0754090 per pixel in the X and Y directions,
respectively for the measurement nearest the date of observation. The
difference in the X and Y plate scales is due to a small tilt of the
IR array relative to the focal plane.  We adopt a zero point of
$M_{H}=21.79\pm0.02$ for F160W for infinite apertures, based on a
comparison of HST archival and ground-based observations of the
standard star P330E (Persson et~al.\ 1998).  It agrees closely with
the value of 21.83 mag obtained by the Space Telescope Science Institute
\footnote{http://www.stsci.edu/ftp/instrument\_news/NICMOS/NICMOS\_phot/keywords.html}.
The foreground Galactic extinction in the direction of Q~1634+267 is only
$E(B-V)=0.072$ for $R_V=3.1$ (Schlegel, Finkbeiner \& Davis 1998); hence, we
applied no extinction corrections to our photometric estimates.

We fitted a point-spread function (PSF) to our combined image, to determine
the relative coordinates and the magnitudes of Q~1634+267 A and B.  Because
A and B are well separated, we derived a PSF estimate from both components
with the IRAF\footnote{IRAF(Image Reduction and Analysis Facility) is
distributed by the National Optical Astronomy Observatories, which are
operated by the Association of Universities for Research in Astronomy, Inc.,
under cooperative agreement with the National Science Foundation.} DAOPHOT
package. Table 2 shows the resulting relative coordinates and H-band
magnitudes of A, B and the star within our field of view.  The quasar images
are separated by 3\farcs852$\pm$0.003; the brightness ratio A/B is $3.34 \pm
0.06$.

We obtained 10-minute snapshots of three quasar pairs, Q~1634+267,
Q~1120+0195, and LBQS~1429--008, on 9 Dec 1998 with the VLA in the C array at
$\lambda = 3.6$ cm.  We detected no radio flux to $1\sigma$ limits of
$0.028$, $0.030$, and $0.045$~mJy/Beam, respectively.  With these low flux
limits, there is little hope that further radio observations can shed much
light on these systems.  

\subsection{Magnitude Limit at the Expected Lens Position}

Visual inspection of the images revealed no lens candidate near the expected
position. To search for a faint lens that might be hidden by A or B, we
fitted the PSF derived in the previous section to the quasar components. We
then subtracted the fitted PSF from these components. The fainter one, B,
was marginally broader than A.  The difference might be caused by intrinsic
PSF differences at different positions on the detector, and similar small PSF
mismatches were seen in other NICMOS archival images containing two or
more stars.  Thus, we cannot conclude that the slight broadening is due to a
faint galaxy underneath quasar B. Such a galaxy would have $m_H\gtrsim21.8$
mag, which is too faint and at the wrong location to be able to produce
lensed images with the observed splitting of $\sim 3\farcs8$.

To determine our detection limit for a lens galaxy, we generated circular
galaxies with a de Vaucouleurs profile and an effective radius corresponding
to an $L^*$ galaxy ($\sim 3h_{65}^{-1}$~kpc) over the redshift range
$0.4\lesssim z\lesssim1.8$ and for $\Omega_0=0.1$ and $1.0$ cosmologies.  We
convolved the redshifted galaxies with the PSF derived from A and B, and
added them to our NIC2 frame on a grid of positions within the frame.  The
resulting images were smoothed with a Gaussian kernel with $\sigma = 3$
pixels, and inspected for signs of the added galaxies.  We found a magnitude
limit of $m_H = 22.5$ ($1\sigma$) for an early-type lens galaxy.  For the
sky background in our frame, the standard deviation of the noise level per
pixel corresponds to a surface brightness of roughly 23.3 mag arcsec$^{-2}$.

\section{LOWER LIMIT ON A LENS M/L RATIO}

We compute the expected mass to light ratio for a putative lens galaxy for
Q~1634+267 from our detection threshold.  The anticipated mass of the lens
can be derived from simple lens theory assuming it has a singular isothermal
sphere (SIS) mass profile.  The lens equation relates the angular
coordinates $\beta$ ($\theta$) of the source (images),
$\vec{\beta}=\vec{\theta} - \vec{\alpha}(\theta)$, where $\alpha(\theta)$ is
the deflection angle for an image at $\theta$, with the lens galaxy at the
center of coordinates. The SIS assumption implies:  $\alpha(\theta) =
(1/2)\Delta\theta = 4 \pi (D_{ds}/ D_s) ({\sigma_v}^2/c^2)$, where $D_s$
($D_{ds}$) is the distance from the observer (lens) to the source;
$\Delta\theta$ is the observed angular separation of the lensed images, and
$\sigma_v$ is the velocity dispersion of the lens galaxy.  Therefore, using
the observed image separation one can determine the velocity dispersion
$\sigma_v$ for a lens galaxy as a function of redshift.  We show in Figure
2a the velocity dispersion ${\sigma_v}(z)$ using cosmologies $\Omega_0=0.1$
and 1, and $H_0=65~\hbox{km~s}^{-1}~\hbox{Mpc}^{-1}$. For an A/B magnification
ratio of $\mu$, the lens galaxy should lie $\Delta\theta\{\mu/(\mu+1)\}$ from
image A along the A-B separation vector. 

We translate our magnitude limit of $H=22.5$ into an upper bound on the
luminosity of the missing lens galaxy, placed at various redshifts.  We
follow the method implemented in Keeton et al.\ (1998) to derive $H$--band
luminosity of elliptical galaxies at various redshifts using
\markcite{charlot1993}Charlot \& Bruzual (1993) models.  The models account
for k-correction and passive evolution from formation redshift $z=5$.
Coupled with the masses required to produce lensing, determined via the lens
equations, the luminosity limits yield the mass-to-light ratio (M/L) as a
function of redshift, shown in Figure 2b for cosmologies of $\Omega_0=0.1$
and $1$.  The most likely (comoving) distance to the lens is half the
distance from the observer to the source, where the galaxy has the largest
cross section for multiply imaging a source.  For Q~1634+267 at redshift
1.961 the most probable lens location is therefore at redshift $0.45$.  An
elliptical galaxy at $z=0.5$ would have $M/L\gtrsim1000$ in a low density
universe, which is $\gtrsim50$ times higher than a normal lens galaxy
(\markcite{keeton1998}Keeton et al.\ 1998,
\markcite{jackson1998}Jackson et al.\ 1998), and 50 times higher than a
limit placed by \markcite{turner1988}Turner et al.\ (1988) in the optical.
Therefore, if obscuration is responsible for hiding a lens galaxy with a
normal mass to light ratio from view, then the dust must cause at least 4
magnitudes of extinction in the $H$-band.  If, however, we accept the
Mg~{\sc ii} absorption lines as evidence for a lens galaxy at $z\approx1.1$
(\markcite{steidel1991}SS91), then the detection limit of $H>22.5$
corresponds to about $0.15L^*_H$, or $M/L_H\gtrsim 690$ for an E/S0 type
galaxy.  But if the lens galaxy type is an Sb, at a redshift of $z=1.1$ it
would have a $M/L_H\gtrsim 430$.  The minimum $M/L_H$ is $400$ for an Sb
galaxy at a redshift of $z=1.3$.

If the lens is an early-type galaxy, we can predict its luminosity from the
Faber-Jackson relation \markcite{faber1976}(1976) between velocity dispersion
$\sigma_v$ and luminosity, ${\sigma_v /\sigma^*}=(L/L^*)^{1/4}$, where
$\sigma^*$ and $L^*$ are the velocity dispersion and luminosity of an $L^*$
galaxy.  Studies of lens statistics (e.g. \markcite{kochanek1996}Kochanek
1996) show that the velocity dispersion estimated from the image separation
should closely match the central velocity dispersion of the stars.  Figure
2a shows the velocity dispersion estimated from the lens geometry as a
function of the lens redshift. At $z\approx0.5$ the predicted velocity
dispersion is $\sigma\approx350~\hbox{km~s}^{-1}$, considerably higher than
the characteristic velocity dispersion of an $L^*$ galaxy of $\sigma^*
\simeq 220\hbox{km~s}^{-1}$.  A galaxy at $z\approx0.5$ would be a
$L\approx6L^*_H$ galaxy, far brighter than normal early-type galaxies,
where $1L^*_H$ would correspond to $H\approx17.4$ $(16.9)$ mag for a
passively evolving stellar population formed at $z_f\approx5$ using the
\markcite{charlot1993}Charlot \& Bruzual (1993) models for $\Omega=0.1$
$(1.0)$.  Such a bright galaxy would be at least $650$ $(1000)$ times above
our detection limit.  At the redshift of the Mg~{\sc ii} absorber of
$z\approx1.1$, the required velocity dispersion rises to
$\approx500~\hbox{km~s}^{-1}$, at the highest end of the range observed for
bright galaxies (see for example, \markcite{bender1996}Bender et al.\
1996).  It would have an extreme luminosity of $15L^*_H$ if the scaling
relation continued to hold.

We conclude that the absolute lower limit on the mass-to-light ratio of the
putative lens is $M/L_H\approx690$ for an early type galaxy at a redshift
$z\approx1.1$, and $M/L_H\gtrsim 400$ for an Sb type at a redshift of
$z\approx1.3$.  The limit is higher for any other redshift and for
$\Omega_0>0.1$.  For the most probable lens redshift of $z\approx0.5$ and
the cosmology of a high-density universe, the limit is $M/L_H>1000$.
Depending on the redshift of the presumed galaxy, the observed $H$ band
corresponds roughly to rest $R$ band ($z\approx1.5$) or $I$ band
($z\approx0.4$).  Since $(B-H)\approx5.7$ mag for ellipticals at redshift
0.5, and $(B-H)_\odot=1.91$ mag, the limit would be about 30 times higher
for $M/L_B$ than for $M/L_H$.  If Q~1634+267A,B is the split image of a
single quasar, the lens must be a galaxy with an unprecedentedly high
mass-to-light ratio.  

In fact, as we discuss in \S4, there are no galaxies above a detection limit
of $\approx 0.1L^*_H$ within $40h_{65}^{-1}$~kpc and no galaxies above a
detection limit of $\approx 1L^*_I$ within $100h_{65}^{-1}$~kpc of the
quasars.  Thus the depth of the imaging also rules out any normal cluster of
galaxies centered near the expected lens position.  Moreover, the limit on
the optical $M/L$ of $M/L_B \approx 10^4h_{65}$ greatly exceeds even the
usual estimates for groups and clusters of galaxies, where virial $M/L
\approx 200$--$400h$ (e.g., see \markcite{carlberg1996}Carlberg et al.\
1996, \markcite{bahcall1995}Bahcall et al.\ 1995, \markcite{david1995}David
et al.\ 1995) is typical.

\section{THE NATURE OF THE Mg~{\sc ii} ABSORBER}

In Q~1634+267A,B, metal absorption lines are observed at $z=1.839$ and
$z=1.126$ (\markcite{steidel1991}Sargent \& Steidel (1991),
\markcite{djorgovski1984}Djorgovski \& Spinrad (1984), and
\markcite{turner1988}Turner et al.\ (1988)).  Mg~{\sc ii}
$\lambda\lambda2796, 2803$ absorption doublets with equivalent widths like
the EW$_0$(Mg~{\sc ii} $\lambda2796)=1.51$\AA\ doublet at $z=1.126$
(\markcite{steidel1991}SS91) are associated with chemically enriched
material in the halos of visible, luminous galaxies along the line of sight
(e.g.\ \markcite{steidel1993}Steidel 1993, \markcite{steidel1995}1995;
\markcite{steidel1997}Steidel et al.\ 1997).  Because quasars show roughly
one metal line absorption system per unit redshift down to a typical rest
equivalent width limit of $\sim0.3$\AA\ \markcite{steidel1993}(e.g. Steidel
1993), the mere presence of a metal absorption system is not evidence for the
presence of a lens galaxy.  In a survey of galaxies selected by the presence
of Mg~{\sc ii} absorption, \markcite{steidel1993}Steidel (1993) showed that
the absorbers are associated with galaxies of Hubble types from E to Sb.
The strength of the absorption is anti-correlated with the distance of the
quasar from the nearest luminous galaxy, and it is correlated with the
intrinsic brightness of the galaxy (\markcite{steidel1995}Steidel
1995,\markcite{steidel1993} 1993).  The luminosities of the $\sim 70$
systems studied by \markcite{steidel1995}Steidel (1995) and
\markcite{steidel1997}Steidel et al.\ (1997) range from $0.02L^*$ to
$4L^*$ with 90\% (61/70) having $L > 0.1L^*$.  All but one (i.e. 99\%)
of the galaxies were found within a projected radius of $40h_{65}^{-1}$~kpc.

We searched both the NICMOS H image and existing HST WFPC1 optical images
(obtained under GTO-1116 by Westphal) for signs of a galaxy associated with
the $z=1.126$ absorption feature.  Since the absorbing galaxy could have an
exponential disk profile rather than the de Vaucouleurs profile considered
in \S2, we re-derived the detection limit for an exponential disk galaxy
with a scale factor of $4\ h^{-1}_{65}$~kpc.  The limit of $22.5$ mag
($1\sigma$ for $\Omega_0=0.1$) was similar to that for a de Vaucouleurs
model with a $3\ h^{-1}_{65}$~kpc effective radius.  In the evolutionary
models of \markcite{charlot1993}Charlot \& Bruzual (1993), a passively
evolving $L^*$, Sb-type, galaxy which formed at $z_f=5$ would have
$H_*\approx19.7$ for $\Omega_0=1.0$, and $H_*\approx20.0$ for
$\Omega_0=0.1$, at $z=1.126$.  Therefore, our detection limit implies that
we could have detected a galaxy with $L > 0.1L^*_H$ anywhere within an
impact parameter at $40h_{65}^{-1}$~kpc from Q~1634+267A (to the edge of the
NICMOS array along the short direction).  We derive a weaker detection limit
of $L>L^*_I$ from the F785LP (I-band) out to $100h_{65}^{-1}$~kpc.  Thus,
any galaxy associated with the $z=1.126$ Mg~{\sc ii} absorption feature is
either unusually faint or distant compared to those found in the
\markcite{steidel1995}\markcite{steidel1993} Steidel (1993, 1995) surveys.

Since the PSF of quasar B is broader than that of quasar A, it is unlikely
that the null detection could be explained by a galaxy directly under
Q~1634+267A, although we cannot exclude a very sub-luminous, compact dwarf
with perfect alignment.  Alternatively, a galaxy with a luminosity $\gtrsim$
$1L^*_B$ lying outside our field of view might be responsible for the
absorption.  Distant ($> 40\ h^{-1}_{65}$~kpc), luminous ($L>L^*$) galaxies
are associated with absorption, but the Mg~{\sc ii} equivalent widths are
less than 0.4\AA\ (\markcite{steidel1995}Steidel 1995,
\markcite{steidel1997}Steidel et al.\ 1997), compared to the observed
1.5\AA.  For distant absorbers it is also odd that no equivalent Mg~{\sc ii}
absorption feature is seen in Q~1634+267B.  In summary, the absorbing galaxy
falls outside of the parameter space populated by galaxies found in Mg~{\sc
ii} absorber survey.  The absorber must therefore be very under-luminous,
$\lesssim 0.1L^*_H$.

\section {SIMILARITY OF THE QUASAR SPECTRA}

In the absence of a lens galaxy, it is the remarkable spectral similarity of
many of the quasar pairs which provides the impetus for the gravitational
lens interpretation.  The spectra of the Q~1634+267A,B pair show some of the
greatest similarities of the pair population, and \markcite{small1997}Small
et al.~(1997) have demonstrated that the observed differences are consistent
with the temporal variations of quasar spectra over epochs separated by
months to years.  No study, however, has examined whether the spectra of the
close pairs show a degree of spectral similarity that is statistically
greater than that of randomly selected quasars.  An affirmative answer
increases the credibility of the lens hypothesis, because the lens hypothesis
provides a natural explanation for the spectral similarities.  While it
would not rule out the binary hypothesis, it leaves a serious puzzle as to
why the nuclear regions of two separate galaxies have such highly correlated
emission properties.  We start by defining how we will measure spectral
similarities, and then estimate the likelihood of finding a single quasar
pair with spectra as similar as Q~1634+267A,B.  In isolation, the significance
is hard to interpret because we picked Q~1634+267A,B for study due to its
spectral similarities.  Thus, in our final step we estimate the likelihood
of finding a population of quasar pairs with the observed spectral
similarities.

\subsection {Sample Definition}

We will compare spectra using the observed differences between the
spectral slope $\alpha$ and the rest equivalent widths (EW$_0$) of the
Ly~$\alpha$, {\sc C~iv} ($\lambda\lambda$1548, 1550) and {\sc C~iii]}
($\lambda$1909) emission lines.  We selected these features because
they are commonly measured in statistical surveys of quasar spectra.
For Q~1634+267 we computed the indices from the
\markcite{steidel1991}SS91 spectra.  These calibrated spectra were
taken at the Palomar 5-meter, using a double spectrograph that covered
the blue (3150--4750\AA) and the red (4600--7000\AA) wavelengths.
We used either the Large Bright Quasar Survey (LBQS) or the pair
population as a whole for a comparison sample.

The LBQS consists of 1055 optically selected quasars
(\markcite{foltz1987}Foltz et al.\ 1987, \markcite{foltz1989} Foltz et al.\
1989, \markcite{hewett1991}Hewett et al.\ 1991).  We use the measurements of
the spectral indices for the LBQS quasars from
\markcite{francis1996}Francis (1996), \markcite{hewett1995}Hewett et al.\
(1995), \markcite{francis1993a}Francis (1993a), and
\markcite{francis1991}Francis et al. (1991).  The binned distributions
for the LBQS quasars were fitted with analytic functions (composites of
Gaussians and Lorentzians) to provide smooth probability distributions.
Some of the spectral features are correlated
(\markcite{francis1992}Francis et al.\ 1992;
\markcite{francis1993a}Francis 1993a), and we will conduct tests to
examine the effects of these correlations on the probability
calculations.

Unfortunately we lack a complete spectral database for the pairs, although
the uncanny similarity of Q~1634+267A,B is not an isolated case.  There are
14 quasar pairs with angular separations
$3\arcsec\lesssim\theta\lesssim10\arcsec$ and similar redshifts (see Table
1).  We will frequently refer to this as our {\it 14 pair sample}.  Of these
14, we can reject 7 as lenses.  In the cases of PKS~1145--071, HS~1216+5032,
Q~1343+2640, J~1643+3156 and MGC~2214+3550, only one of the quasars is a
radio source; in the case of Q~0151+048 the small emission line redshift
difference is confirmed by an absorption feature in the background quasar;
in LBQS~2153--2056, the spectrum of quasar B shows a strong absorption
blueward of the {\sc C~iv} emission line.  Four of these binaries
(QJ~0240--343, PKS~1141--071, HS~1216+5032, Q~0151+048) show fairly similar
spectra, although they all have at least one significant spectral
difference.  Of the 7 ambiguous pairs, five (Q~2138--431 in
\markcite{hawkins1997}Hawkins et al.\ 1997, Q~1429--008 in
\markcite{hewett1989}Hewett et al.\ 1989, Q~1634+267 and Q~2345+007 in
\markcite{steidel1991}SS91, Q~1120+0195 in
\markcite{michalitsianos1997}Michalitsianos et al.\ 1997) show
remarkably similar optical spectra.  For the pair sample we restrict
our comparison to the {\sc C~iv} equivalent width and the power law
slope $\alpha$ because the other two indices are not consistently
observed or are impossible to measure without the original data.

\subsubsection {Line and Continuum Slope Measurements}

To determine the spectral slope, we fit a power law (defined as
F$_\nu\propto\nu^\alpha$) excluding regions where there are emission or
absorption lines.  The Fe~{\sc ii} in the UV spectra of quasars introduce
substantial uncertainty in the measurement of continuum slopes.  The power
law $\alpha$ is also uncertain in the blue spectrum from contamination due
to the extended wings of the {\sc C~iv}, {\sc N~v}, and Ly~$\alpha$ emission
lines, as well as the Ly~$\alpha$ forest absorption.  Because of these
complications, the width of the spectral slope distribution remains
controversial.  Some (e.g. \markcite{elvis1994}Elvis et al.\ 1994;
\markcite{webster1995}Webster et al.\ 1995) find large dispersions,
while others find little or no dispersion in $\alpha$
(\markcite{francis1996}Francis 1996; \markcite{sanders1989} Sanders et al.\
1989).  In our analysis we will adopt the narrow $\alpha$ distribution from 
\markcite{francis1996}Francis (1996), as derived from 30 LBQS quasars.  The
Francis (1996) distribution has a root-mean-squared (RMS) width of
$\sigma_\alpha=0.3$, compared to $\sigma_\alpha=0.5$ for the broader
distributions of \markcite{francis1992}Francis et al. (1992).  In Q~1634+267
the mean slopes are $\alpha=-0.17$ in the blue (observed 3150--4734\AA) and
$\alpha=-0.12$ in the red (observed 4580--6985\AA) spectra respectively.  In
both the blue and the red, the power law indices between A and B differ by
$\Delta\alpha=0.03\pm0.03$.

When comparing equivalent widths we must be careful to use identical
measurement methods because of the difficulties introduced by line blending
and continuum definition.  \markcite{francis1993a}Francis (1993a) showed that
{\sc C~iv} equivalent widths can vary by a factor of two depending on the
measurement method, and thus broaden the distribution of equivalent widths
by the same factor.  For Q~1634+267A,B we avoid these problems by using the
same definitions followed by the LBQS (\markcite{francis1993a}Francis 1993a).
For this method, continuum windows are defined on both sides of a given
line.  Then the flux is summed above a line adjoining the continuum windows,
over a region known as the line window.  The continuum windows are chosen to
exclude the flux contributed by nearby broad lines, so the line fluxes
should be considered a lower limit.  The difference in rest equivalent widths
(EW$_0$) between A and B are 3\AA\ for {\sc C~iv}, 8\AA\ for Ly~$\alpha$
and 3\AA\ for {\sc C~iii]}.  Measuring EW$_0$({\sc C~iv}) requires
connecting the red and blue halves of the spectra because the blue side
spectrum terminates on the red wing of the {\sc C~iv} emission line, causing
the continuum windows to fall on different spectrographs.  We calculate a
scaling factor of 1.22 using the regions of overlap.  But there is some
evidence that the same scaling factor might not apply to quasar B because of
large fluctuations in the data over that region.  Therefore the EW$_0$({\sc
C~iv}) differences of 3\AA\ should be considered an upper limit.  Indeed,
restricting ourselves to the blue regions alone and fitting Gaussians to the
emission lines we find quasars A and B to differ at most by
$\Delta$EW$_0$({\sc C~iv})$~=1$\AA.  The two methods yield the same flux.
For completeness, and because the numbers have not been published elsewhere,
we also measure other line fluxes and summarize them in Table 2, although
some we do not use in our analysis.  All of the line fluxes are measured with
the technique described in \markcite{francis1993a}Francis (1993a) except for
{\sc N~v}.  Because the blending with Ly~$\alpha$, the measurement for {\sc
N~v} lines is done by fitting Gaussians to Ly~$\alpha$ simultaneously with
{\sc N~v} where we assumed that both lines are similar in shape.

For the other quasar pairs we estimated bounds on the {\sc C~iv} equivalent
width and the spectral index $\alpha$ based on the published spectra.  For
the 5 most similar pairs, we estimate that they have spectral differences
$\Delta\alpha\lesssim0.1$ and $\Delta$EW$_0$({\sc C~iv})~$\lesssim6$\AA\
(see Table 3).  For the two pairs whose spectra have the worst noise
(Q~1120+0195 (\markcite{michalitsianos1997}Michalitsianos et al.\ 1997) and
Q~1429--008 (\markcite{hewett1989}Hewett et al.\ 1989)) the estimates are
rough, while they are more accurate for the other three pairs.  Since the
estimates of the equivalent widths may be imprecise, we will also quote
results by doubling our bound on the differences to $\Delta$EW$_0$({\sc
C~iv})~$\lesssim12$\AA.  We summarize our parameter estimates for the pairs
in Table 3.

\subsubsection {Joint Probabilities and Correlations among Spectral Features}

To compute the probability that two random quasars have given differences in
{\sc C~iii]}, Ly~$\alpha$, {\sc C~iv} equivalent widths and spectral slope
$\alpha$, we randomly draw the properties of two quasars at a time from the
compiled number distributions.  If the spectral features are completely
uncorrelated, the likelihood that two quasars are similar to each other, i.e.
have differences of less than $\Delta$EW$_0$({\sc C~iv}), $\Delta$EW$_0$({\sc
C~iii]}), etc., is simply the joint product of each individual probability.
However, if there are correlations, they could make the alarming
similarities of the quasar pairs a trivial selection effect.  Therefore we
first explore the extent to which spectral features are correlated.

Through a principal component analysis, \markcite{francis1992}Francis et
al.\ (1992) showed that the equivalent widths of quasar emission lines are
correlated with the power law slope $\alpha$. \markcite{francis1993b}Francis
(1993b) further showed that $\alpha$ is correlated with redshift.  Their
study was based on correlations with a spectral slope distribution in
\markcite{francis1992}Francis et al.\ (1992) which has a large RMS
dispersion of about $\sigma_\alpha=0.5$.  The strongest correlation, Al~{\sc
iii}$+${\sc C~iii]} with $\alpha$, has a large scatter; the correlations of
this and other lines are very weak.  Subsequently,
\markcite{francis1996}Francis (1996), using more accurate photometry, found
that the spectral slope distribution actually has a much narrower dispersion
of $\sigma_\alpha=0.3$.  But the newer study does not affect the dispersions
of the line strengths, which are more accurate as long as they are measured
consistently.  Because the scatter of the line strengths remains large for a
much reduced range in spectral slopes, the correlations are weaker in light
of the narrow distribution compared to \markcite{francis1992}Francis et al.\
(1992).  Furthermore, plotting spectral slope {\it vs.} redshift from
\markcite{francis1996}Francis (1996) shows that there is no correlation
between the two parameters.  In the analysis below, we will use the narrow
power law distribution found in \markcite{francis1996}Francis (1996).
Therefore, the correlations between line emission, continuum, slopes and
redshifts can be safely ignored for our purposes.

Nonetheless, to confirm that correlations are unimportant even under
extreme scenarios we conduct the following Monte Carlo experiment to choose
quasars from the number distributions found in \markcite{francis1992}Francis
et al.\ (1992) and \markcite{francis1993a}Francis (1993a).  We select from the
broader spectral slope distribution where correlations were originally
found.  The correlation between emission line strengths and $\alpha$,
and $\alpha$ with redshift $z$, means that the process of
drawing quasars must proceed in a sequence to account for
correlations at each step:  first we draw quasars out of the LBQS redshift
distribution, then the spectral slope distribution, followed at last by the
emission line distribution.  To account for the correlation of $\alpha$ with
redshift we do the following.  Having picked a redshift, and given the power
law distribution $\alpha$, we offset the entire distribution by an amount
$d\alpha=0.79\times d~log(1+z)$ (See Figure 2 of
\markcite{francis1993b}Francis 1993b).  We can then randomly select a value
of $\alpha$ from this new distribution.  There is, however, one subtlety.
The distribution for $\alpha$ contains no redshift information because it is
summed over all redshifts.  But because of the correlation, if quasars at
{\it any given} redshift have a small dispersion in their spectral slopes,
as one accumulates the distribution over broader redshift ranges, the {\it
net} distribution would also gradually become wider.  We account for this
effect by arbitrarily shrinking the dispersion and then drawing from this
narrower spectral slope distribution instead of the original, broader one.
We verified that decreasing the dispersion by 25\%, more than one would
expect, would not significantly alter our discussions below.  Once a value
for $\alpha$ has been selected, one can then proceed in the same fashion to
select from the number distribution of {\sc C~iii]}, {\sc C~iv}, or
Ly~$\alpha$, accounting for their roughly linear correlations with $\alpha$
by:  $EW_{line}=m\alpha+b$ where the correlation coefficients $m$ and $b$
are given in \markcite{francis1992}Francis et al.  (1992).  Again, we
verified that any correlations would only begin to affect our conclusions if
the number distributions were all narrowed by more than 25\%.

There may also be weak correlations of quasar spectral properties with
luminosity. The pair members share a common redshift and have similar
luminosities, so we would under-estimate the likelihood of the members
possessing similar spectra if the distributions of spectral indices are
correlated with either variable.  We saw that the spectral index correlation
with redshift has little or no role in our analysis.  The Baldwin
(\markcite{baldwin1977}Baldwin 1977) effect, a correlation between
luminosity and {\sc C~iv} equivalent width, has been studied extensively
(e.g. more recently see \markcite{kinney1990}Kinney et al.\ 1990,
\markcite{sargent1989}Sargent et al.\ 1989,
\markcite{baldwin1989}Baldwin et al.\ 1989).  Their results show that at a
given luminosity the equivalent widths of the {\sc C~iv} emission line have
a defined range in strengths, with a scatter of about 20\AA.  The strength
of the {\sc C~iv} lines decreases with luminosity and the relation holds
over 7 orders of magnitude in quasar luminosity.  The parameterization of
the correlation: $EW({\sc C~iv})\propto L^{\beta}_{1450}$ has $\beta$
ranging from $\beta=-0.07$ (\markcite{francis1992}Francis et al.\ 1992) to
$\beta=-0.45$ (\markcite{smetanka1991}Smetanka et al.\ 1991).  LBQS quasars,
being a magnitude limited sample, have only a small spread in absolute
luminosity, $-28\lesssim M_B\lesssim-26$ \markcite{francis1992}(Francis et
al.\ 1992).  Since at any luminosity there is a considerable scatter in {\sc
C~iv} and vice versa (see \markcite{osmer1998}Osmer \& Shields 1988 for
summary), we do not expect the luminosity correlations to affect our
analysis. Nonetheless, we tested for any effects by deriving the distribution
of $\Delta\hbox{EW}_0({\sc C~iv})$ both with and without limits on the 
luminosity difference $\Delta L$ of the quasars making up the simulated
pairs using the $B_J$ magnitudes and EW$_0$({\sc C~iv}) values for the
439 LBQS survey quasars from Francis (1993).  We found no significant
differences between the distributions without a luminosity restriction
and those where the quasars were required to have 
absolute luminosities differing by less
than a factor of 5 ($1.7$ mag) or a factor of 2.5 ($1$ mag). 

Because the parameter correlations are very weak, in the discussions
that follow, we make the assumption that no two spectral features are
correlated.

\subsection {The Spectral Similarity of Q~1634+267A,B}

As discussed previously, the spectra of Q~1634+267 A and B have equivalent
width differences $\Delta\leq3$\AA\ for ${\sc C~iv}$; 8\AA\ for Ly~$\alpha$;
and 3\AA\ for {\sc C~iii]}, as well as 0.03 for $\alpha$.  If we select
randomly from the distributions of these features in the LBQS survey, the
probability of finding two quasars with smaller differences are 11\%, 16\%,
30\%, and 12\% respectively.  Assuming no correlations between the features,
the joint probability of the differences being as small as observed is
0.06\%.

With such a low joint probability, it would be reasonable to find several
quasars with such similar spectra in the total quasar population ($\sim 10^4$
quasars), but highly unlikely to find any in the tiny sub-population of
quasar pairs ($14$ objects).  The strength of this conclusion depends only
weakly on the details of the comparison, and the joint probability only rises
to 1\% if we confine the comparison to the spectral index and {\sc C~iv}
equivalent width differences.  Adding additional spectral features to the
estimate would further reduce the likelihood, but it becomes increasingly
important to fully understand the spectral correlations of quasars.
Moreover, calculations of {\it a posteriori} probabilities can generally
yield arbitrarily low likelihoods by over-specifying the problem.
Nonetheless, the similarity of the Q~1634+267 A and B spectra extends across
the entire rest-UV spectra of the pair.

\subsection {The Spectral Similarity of the Pair Population}

Because we pre-selected Q~1634+267A,B for study because of its spectral
similarities, we cannot interpret it in isolation from the rest of the pair
population.  Moreover, the uncanny similarities of Q~1634+267A,B are not an
isolated example among the pairs. To fully understand the significance of
the spectral similarities of the pairs, we must consider the entire pair
population.

Under the assumption that the quasars making up the pair sample are drawn
from the same parent population as the quasars in the LBQS, we can produce
mock samples of quasar pairs by randomly selecting 14 pairs of quasars from
the LBQS.  We can then compare the distribution of the spectral
slope and {\sc C~iv} equivalent width differences, $\Delta \alpha$ and 
$\Delta$EW$_0$({\sc C~iv}), to those observed for the pairs.  We will
estimate the probability of finding at least as many similar pairs as
observed in the real sample.  

We generated $10^4$ mock pair catalogs to determine the fraction of catalogs
in which N$_{sim}$ pairs have spectral differences smaller than the
specified limits on $\Delta\alpha$ and $\Delta$EW$_0$({\sc C~iv}), with the
results illustrated in Figure 3.  The top panel shows the limits on spectral
similarity ($\Delta\alpha\lesssim0.1$ and $\Delta$EW$_0$({\sc
C~iv})$\lesssim6$\AA) which characterize the differences observed for the 5
pairs with similar spectra (Q~2138--431, LBQS~1429--008, Q~1634+267,
Q~1120+0195 and Q~2345+007).  The chance of drawing 5 or more pairs with
such similar spectra in the sample of 14 pairs is about 0.2\%.  If we were
to relax the constraints to $\Delta\alpha\lesssim0.1$ and
$\Delta$EW$_0$({\sc C~iv})$\lesssim12$\AA\ to account for possible
systematic errors in measuring $\Delta$EW$_0$({\sc C~iv}), the probability
for finding 5 or more such pairs grows only to 3\%.  We conclude that the
likelihood for finding 5 pairs with such similar spectra in a sample of 14
pairs is low.  If we significantly broaden the criterion for similarity
($\Delta\alpha\lesssim0.2$ and $\Delta$EW$_0$({\sc C~iv})$\lesssim20$\AA),
the likelihood of finding five or more similar pairs rises to 60\%.
However, with these weaker criteria there are now seven observed pairs
meeting the criterion for similarity (we should add QJ~0240--343 and
LBQS~2153--2056), and the likelihood of finding seven or more similar pairs
is then smaller, at 20\%.  We regard this third case as a significant
under-estimate of the similarities.  The low probability for finding similar
quasars randomly, using two parameters, may only be an upper bound.  If our
pairs sample were partly made up of gravitational lenses, microlensing and
time delay might cause some spectral differences in those systems.  It is
impossible to quantify such a conjecture with our pairs sample.

The probability for finding similar quasars is significantly lower if we use
a broad continuum slope distribution (e.g. \markcite{francis1992}Francis
1992) rather than the narrow distribution of \markcite{francis1996}Francis
(1996) used for the estimates in Figure 3.  For the broad estimate of the
$\alpha$ distribution, the likelihoods for drawing at least 5 similar pairs
under the same three sets of constraints on $\Delta\alpha$ and
$\Delta$EW$_0$({\sc C~iv}) are 0.01, 0.2\% and 1.2\% instead of 0.2\%, 3.0\%
and 20\% (see Figure 4).  Thus, the debate about the width of the
continuum slope distribution does not alter the conclusion that the spectral
similarities of the pairs are unlikely to occur by chance unless the
distribution is even narrower than that found by
\markcite{francis1996}Francis (1996).  The distribution of {\sc C~iv}
equivalent widths, {\it as long as measured consistently}, does not suffer
from the systematic problems of the spectral slope distribution.  

\section{CONCLUSIONS}

We again confirm that there is no sign of a lensing galaxy or a surrounding
group or cluster of galaxies near the Q~1634+267A,B quasar pair.  Our limit
on the mass-to-light ratio is $\gtrsim690h_{65}$ in the $H$ band, tighter than
previous estimates.  We also lack a candidate galaxy within
$100h_{65}^{-1}$~kpc for producing the $z=1.126$ Mg~{\sc ii} absorption
feature in the spectrum of A.  Since the existence of even one large
separation lens produced by a ``dark'' mass distribution has profound
implications, confirming or disproving the lens hypothesis is highly
desirable.  It appears from the data in the literature that the flux ratio
of the quasars is significantly time variable, changing from 4.4
(\markcite{djorgovski1984}Djorgovski \& Spinrad 1984) to 2.83
(\markcite{turner1988}Turner et al.\ 1988), to 3.28
\markcite{steidel1991}(SS91) and 3.34 (this work).  With its large angular
separation, regular photometric monitoring of the system to search for a
time delay would be simple on modest size, ground-based telescopes.
Although we found that the pairs Q~1634+267, Q~1120+0195, and LBQS~1429--008,
are at best weak radio sources, with upper limits on their fluxes of $0.28$,
$0.030$, and $0.045$~mJy/Beam at $\lambda=3.6$ cm, deeper observations may
still detect the sources or demonstrate that both members are at the
extremes of the radio to optical flux ratio distribution.  Deep X-ray
observations would also be useful, both to measure the flux ratio of the
quasars and to search for hot X-ray emitting gas associated with any
lensing potential.  Deep IR imaging should detect the quasar host galaxies,
whose shapes must show tangential, arc distortions under the lens hypothesis.

The spectral similarities of some of the quasar pairs has always been the
major impetus for the ``dark'' lens interpretation of the systems. To date,
these arguments have been based on the similarities of individual pairs
rather than the overall pair population.  Using the distribution of
continuum slopes and {\sc C~iv} equivalent widths, we demonstrated that it
was improbable to find as many similar pairs as observed in the existing
sample of 14 pairs.  The probability estimates range from 0.01\% to 3\%,
depending on how strictly similarity is defined and depending on whether a
broad or narrow distribution was used for the continuum slope distribution.
A better quantitative analysis needs a uniform spectral survey of the pairs. 

The effects of binary quasar interaction on spectral similarity have been
surmised, for example, by \markcite{schneider1993}Schneider (1993).
\markcite{kochanek1999}Kochanek et al. (1999) demonstrated that most of the
quasar pairs are binary quasars rather than ``dark lenses.'' The binary
quasar hypothesis is a successful quantitative interpretation of both the
number of quasar pairs and the distribution of their separations.  If they
are all binary quasars, then there must be a physical explanation for the
spectral similarities.  Binary quasars are examples of quasar activity
triggered by mergers, which represent a small fraction of all quasar
activity (about 5\% in simple models, see Kochanek et al. 1999a).  We
present two hypotheses for how binary quasars can have such similar
spectra.  First, if the physical properties of quasars triggered by major
mergers (e.g. mass, accretion rate, surrounding gas density) were confined
to a limited range of values compared to the overall quasar population, then
the observable properties of the binaries would be more similar than those
of randomly selected quasars.  This hypothesis has a definite, testable
prediction -- the spectra of the different pairs should be more similar to
each other than for randomly selected quasars.  It would also be expected
that the quasar host galaxies have properties commensurate with being the
products of major mergers -- which can be tested by deep imaging.  Second,
if quasar spectra have a significant dependence on the age of the activity,
then the binary spectra are more similar than for random quasars because the
merger provides a common triggering event for the activity in the two
systems.  Here we would not expect the spectra of the different pairs to
show any more similarity than randomly selected quasars.  In either case,
the binary quasar population becomes a very important probe of the physics
of quasars because it provides evidence for correlations between events on
large scales and the detailed operation of the central engine.

\section{ACKNOWLEDGMENTS}

We are very grateful to C.C. Steidel for providing the Q~1634+267 spectrum
used in our analysis and to P.J. Francis for supplying the luminosity and
{\sc C~iv} data for the LBQS quasars.  We also thank C.B. Foltz, L.C. Ho,
A.V. Filippenko, W.D. Li, A.  Quillen, and G. Rudnick for enlightening
discussions, and the referee P. Schneider for his comments.  Support for the
CASTLES project was provided by NASA through grant numbers GO-7495 and
GO-7887 from the Space Telescope Science Institute, which is operated by the
Association of Universities for Research in Astronomy, Inc.  CSK and CRK
were also supported by the NASA Astrophysics Theory Program grant
NAG5-4062.  HWR is also supported by Fellowship from the Alfred P. Sloan
Foundation. Our research was supported by the Smithsonian Institution.

\vfill\eject
\clearpage
\begin{deluxetable}{lcccccccccc}
\scriptsize
\tablewidth{0pt}
\tablecaption{Quasar Pairs With Separations of 3\arcsec\ to 10\arcsec}
\tablehead{ Name &$z_s$ &$\Delta \theta$ &$R$                &$f_O$ &$f_R$ &$F_{20}$ &$|\Delta v|$ &Lens? &Type &Ref \\
                 &      &                &$h_{50}^{-1}$ kpc  &      &      &mJy      &km s$^{-1}$       &      &     & }
\startdata
\tableline
 Q~0151+048$^{\bf\dag}$     &$1.91$ &$3\farcs3$ &$28$ & 27.5 &         &$<1$     &$520\pm160$    &No   &$O^2$    &1  \nl
 PKS~1145--071              &$1.35$ &$4\farcs2$ &$36$ &  2.1 &$>500$   &$740$    &$200\pm110$    &No   &$O^2R$   &2  \nl
 HS~1216+5032               &$1.45$ &$9\farcs1$ &$78$ &  5.2 &$>28$    &$3.9$    &$260\pm1000$   &No   &$O^2R$   &3  \nl
 Q~1343+2640                &$2.03$ &$9\farcs5$ &$78$ &  1.1 &$>57$    &$8.6$    &$120\pm890$    &No   &$O^2R$   &4 \nl
 J~1643+3156                &$0.59$ &$2\farcs3$ &$17$ &  1.3 &$>40$    &$120$    &$  80\pm10$    &No   &$O^2R$   &5 \nl
 LBQS~2153--2056            &$1.85$ &$7\farcs8$ &$64$ & 14.5 &         &$<1$     &$1100\pm1500$  &No   &$O^2$    &6 \nl
 MGC~2214+3550              &$0.88$ &$3\farcs0$ &$26$ &  1.6 &$>42$    &$246$    &$148\pm420$    &No   &$O^2R$   &7 \nl
 MG~0023+171                &$0.95$ &$4\farcs8$ &$40$ &  3.0 &$\sim10$ &$186$    &$292\pm260$    &?--  &$O^2R^2$ &8  \nl
 Q~1120+0195$^{\bf\dag\dag}$&$1.46$ &$6\farcs5$ &$56$ &173.8 &         &$<1$     &$628\pm120$    &?--  &$O^2$    &9  \nl
 LBQS~1429--008             &$2.08$ &$5\farcs1$ &$42$ & 17.4 &         &$<1$     &$260\pm300$    &?--  &$O^2$    &10 \nl
 QJ~0240--343               &$1.41$ &$6\farcs1$ &$52$ &  2.1 &         &$<1$     &$250\pm180$    &?    &$O^2$    &11 \nl
 Q~1634+267                 &$1.96$ &$3\farcs8$ &$32$ &  4.4 &         &$<1$     &$33\pm86$      &?    &$O^2$    &12 \nl
 Q~2138--431                &$1.64$ &$4\farcs5$ &$38$ &  3.0 &         &         &$0\pm115$      &?    &$O^2$    &13 \nl
 Q~2345+007                 &$2.15$ &$7\farcs3$ &$58$ &  4.0 &$>9$     &$0.035$  &$476\pm500$    &?    &$O^2$    &14 \nl
 RXJ~0911.4+0551            &$2.80$ &$3\farcs1$ &$24$ &  1.9 &         &$<1$     &$158\pm1000$   &Yes  &$O^2$    &15 \nl
 Q 0957+561                 &$1.41$ &$6\farcs1$ &$52$ &  1.4 & $1.3$   &$552$    &$200\pm15$     &Yes  &$O^2R^2$ &16 \nl
 HE~1104--1805              &$2.32$ &$3\farcs1$ &$24$ &  4.8 &         &$<1$     &$300\pm90$     &Yes  &$O^2$    &17 \nl
 MG~2016+112                &$3.27$ &$3\farcs6$ &$26$ &  1.7 &$\sim1$  &$191$    &$40\pm100$     &Yes  &$O^2R^2$ &18 \nl 
\tableline
\enddata
\tablecomments{ $z_s$ is the source redshift, $\Delta \theta$ is the angular
separation, $R$ is the projected separation at the source redshift for
$\Omega_0=1$ and $H_0=50h_{50}$ km s$^{-1}$ Mpc$^{-1}$, $f_0$ and $f_R$ are
the optical and radio flux ratios or their lower limits, $F_\lambda$ is the
radio flux of the brighter image at wavelength $\lambda$ (in cm), and
$|\Delta v|$ is the velocity difference between the quasars.  The entries in
the Lens? column are: ``Yes'' if a normal lens (galaxy, group, or cluster)
is seen in the correct position to produce the observed system, there is no
significant velocity difference, and the radio and optical data are
consistent with the lens hypothesis; ``No'' if we see no lens and either the
radio emission or the emission line velocity difference, confirmed by an
absorption line velocity difference, are inconsistent with the lens
hypothesis; and, ``?'' if we see no lensing object but have no objective
criterion to decide whether or not the object is lensed.  If there is some
evidence that the system is actually a binary, we used the label ``?--''.
Type denotes the optical/radio classification of the pair.  Note that
MG~2016+112 is really a triple system, not a pair. \\ $^{\bf\dag}$Q~0151+048
is also named PHL~1222 and UM~144.\\ $^{\bf\dag\dag}$Q~1120+019 is also
named UM~425. \\ References: (1) Meylan et al.\ 1990, (2) Djorgovski et al.\
1987, (3) Hagen et al.\ 1996, (4) Crampton et al.\ 1988, (5) Brotherton et
al.\ 1999, (6) Hewett et al.\ 1997, (7) Mu\~noz et al.\ 1997, (8) Hewitt et
al.\ 1987, (9) Meylan \& Djorgovski 1989, (10) Hewett et al.\ 1989, (11)
Tinney 1995, (12) Djorgovski \& Spinrad 1984, (13) Hawkins et al.\ 1997,
(14) Weedman et al.\ 1982, (15) Bade et al.\ 1997, (16) Walsh et al.\ 1979,
(17) Wisotzki et al.\ 1993, (18) Lawrence et al.\ 1984.}
\end{deluxetable}

\begin{deluxetable}{lr}
\tablewidth{0pt}
\tablecaption{Observed Properties of Q~1634+267 A,B}
\tablehead{ \colhead{Parameter} & \colhead{Value} }

\startdata
Image separation                                & $3\farcs852\pm0.003$ \nl
\ \ \ \ $\Delta$ RA                             & $-0\farcs673$ \nl
\ \ \ \ $\Delta$ DEC                            & $3\farcs793$ \nl
H-magnitude of component A                      & $18.26\pm0.01$ \nl
H-magnitude of component B                      & $19.56\pm0.01$ \nl
H-magnitude of star                             & $19.67\pm0.01$ \nl
Flux ratio (A/B)                                & $3.34\pm0.06$ \nl
Power law slope $\alpha$\tablenotemark{a}       & $-0.13$ \nl
EW$_0$(Ly~$\alpha$)                             & $47.95$\AA\nl
EW$_0$({\sc N~v} $\lambda$1240)                 & $24.18$\AA\nl
EW$_0$(Si~{\sc iv} $\lambda$1400)               & $11.83$\AA\nl
EW$_0$({\sc C~iv} $\lambda\lambda$1548, 1550)   & $37.44$\AA\nl
EW$_0$({\sc C~iii]} $\lambda$1909)              & $16.91$\AA\nl
\enddata
\tablenotetext{a}{Power law slope defined as F$_\nu \propto\nu^\alpha$}
\end{deluxetable}

\begin{deluxetable}{lrccl}
\scriptsize
\tablewidth{0pt}
\tablecaption{QSOs A,B Parameter Difference Estimates}
\tablehead{ &\colhead{$\Delta\alpha$} & 
\colhead{$\Delta$EW$_0$(C~{\sc iv}) \AA} & \colhead{Lens?} & 
\colhead{References}}

\startdata
Q~1120+0195       &  0.1  &   6 & ?  &  Michalitsianos et al.\ 1997\nl
LBQS~1429--0053   &  0.03 &   6 & ?  &  Hewett et al.\ 1989\nl
Q~1634+267        &  0.03 &   3 & ?  &  Steidel \& Sargent 1991\nl
Q~2138--431       &  0.02 &   6 & ?  &  Hawkins et al.\ 1997\nl
Q~2345+007        &  0.07 &   3 & ?  &  Steidel \& Sargent 1991\nl
LBQS~2153--2056   &  0.05 &  14 & No &  Hewett et al.\ 1998\nl
QJ~240--343       &  0.2 \tablenotemark{a} &  20 &  No &Tinney 1995\nl 
\enddata
\tablecomments{Estimated differences between the spectra of QSO pairs
A and B from published spectra in literature.}
\tablenotetext{a}{From Tinney 1995.}
\end{deluxetable}

\vfill\eject
\clearpage
\begin{figure}
\plotone{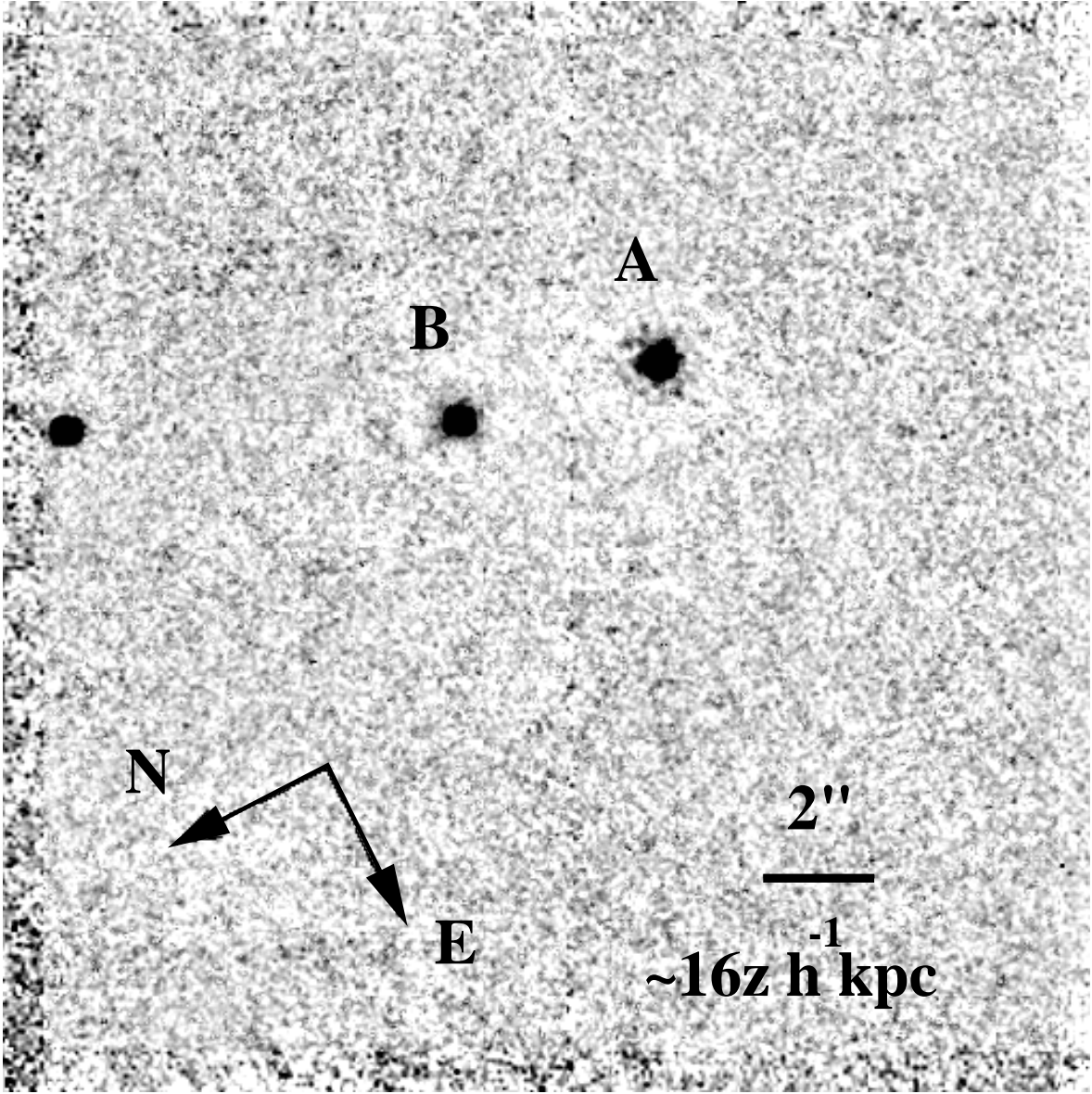}
\caption[Figure 1]{Image of Q~1634+267 A and B.  Horizontal bar shown is
2\arcsec\ in length, which corresponds to roughly
16~$z~h_{65}^{-1}\hbox{kpc}$, for an $\Omega=0.1$ open cosmology.}
\end{figure}

\begin{figure}
\plotone{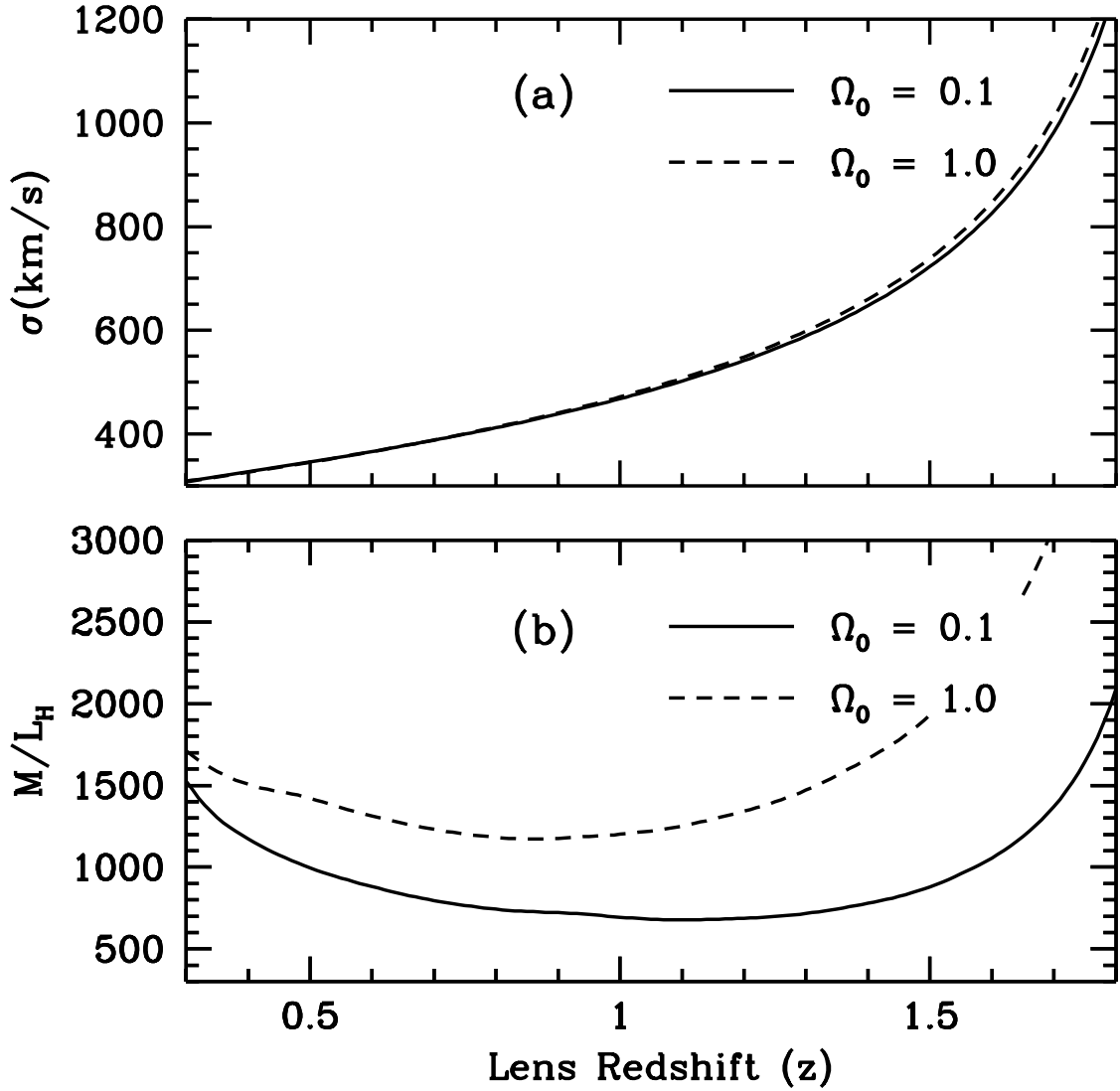}
\caption[Figure 2] {(a) Anticipated velocity dispersion of a putative lens
galaxy and (b) $M/L_H$ ratio as function of redshift for a putative lens
galaxy, assuming properties of an early-type galaxy.}
\end{figure}

\begin{figure}
\plotone{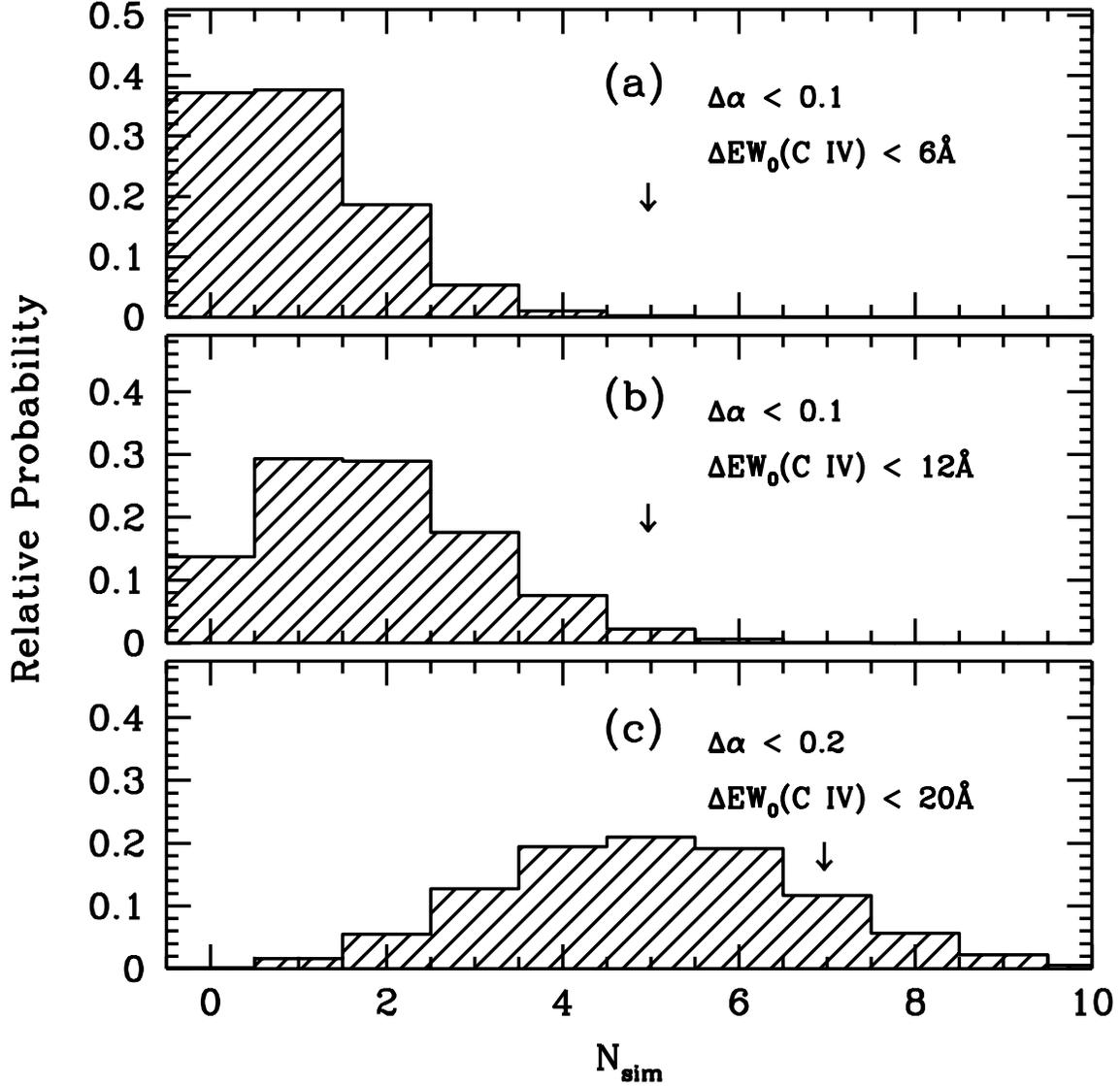}
\caption[Figure 3] {The fraction of trials in which $N_{sim}$ pairs from a
sample of 14 pairs randomly drawn from the LBQS meet the stated conditions
on $\Delta\alpha$ and $\Delta $EW$_0$({\sc C~iv}).  We assume the narrow
spectral slope distribution from \markcite{francis1996}Francis (1996).  The
arrow marks the number of pairs observed to meet the similarity condition.
(a) Representative spectral differences actually observed for the five lens
candidates.  Probabilities for (b) and (c) have substantially weaker {\sc
C~iv} and spectral slope constraints, as indicated in each box.}
\end{figure}

\begin{figure}
\plotone{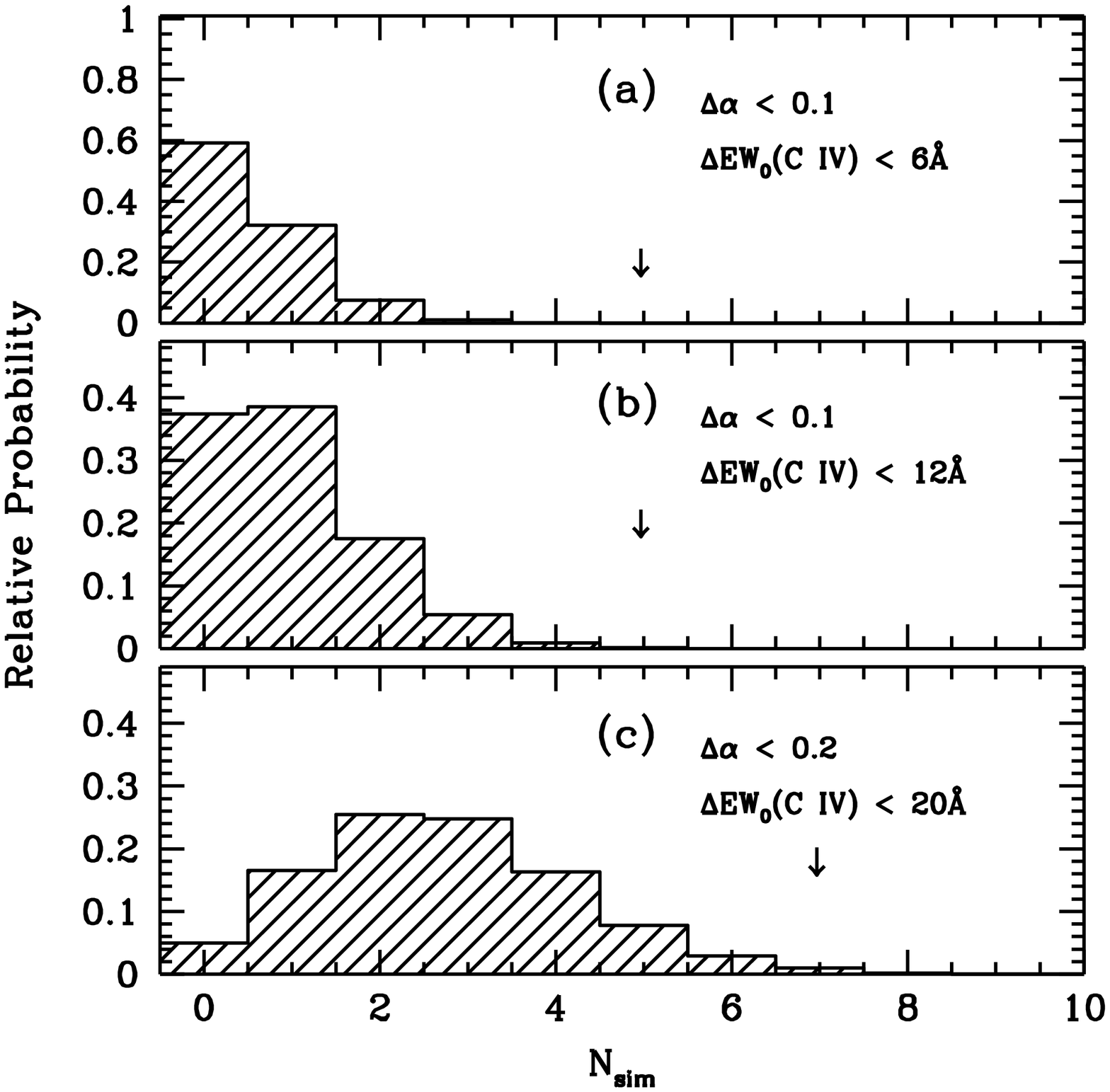}
\caption[Figure 4] {The fraction of trials in which $N_{sim}$ pairs from a
sample of 14 pairs randomly drawn from the LBQS meet the stated conditions
on $\Delta\alpha$ and $\Delta $EW$_0$({\sc C~iv}). Here we assume the
broad spectral slope distribution from \markcite{francis1992}Francis et al.\
(1992). The arrow marks the number of pairs observed to meet the similarity 
condition.  The cases similarity conditions are the same as in Figure 3.}
\end{figure}


\begin{references}

\reference{bade1997} Bade, N., Siebert, J., Lopez, S., Voges, W., Reimers,
D., et al.\ 1997, \aap, 317, 13

\reference{bahcall1995} Bahcall, N. A., Lubin, L. M., \& Dorman, V. 1995,
\apj, 447, L81

\reference{baldwin1977} Baldwin, J. A. 1977, \apj, 214, 679

\reference{baldwin1989} Baldwin, J. A., Wampler, E. J., \& Gaskell, C. M.
1989, \apj, 338, 630

\reference{bender1996} Bender, R., Ziegler, B., Bruzual, G. 1996, \apjl, L51

\reference{brotherton1999} Brotherton, M. S., Gregg, M. D., Becker, R. H.,
Laurent-Muehleisen, S. A., et al.\ 1999, \apjl, 514, L61

\reference{carlberg1996} Carlberg, R. G., Yee, H. K. C., Ellingson, E.,
Abraham, R., et al.\ 1996, \apj, 462, 32

\reference{charlot1993} Charlot, S. \& Bruzual, A. G. 1993, \apj, 405, 538

\reference{cox1997} Cox, C. et al.\ 1997, Instrument Science Report,
OSG-CAL-97-07, Space Telescope Science Institute

\reference{crampton1988} Crampton, D., Cowley, A. P., Hickson, P., Kindl, E.,
et al.\ 1988, \apj, 330, 184

\reference{david1995} David, L. P., Jones, C., \& Forman, W. 1995, \apj,
445, 578

\reference{djorgovski1984} Djorgovski, S., \& Spinrad, H. 1984, \apjl, 282,
L1

\reference{djorgovski1987} Djorgovski, S., Perley, R., Meylan, G., \&
McCarthy, P.  1987, \apjl, 321, L17

\reference{elvis1994} Elvis M., Wilkes, B. J., McDowell, J. C., Green, R. F.,
et al.\ 1994, \apjs, 95, 1

\reference{faber1976} Faber, S. M. \& Jackson, R. E. 1976, \apj, 204, 668

\reference{filippenko1989} Filippenko, A. V. 1989, \apjl, 338, L49

\reference{fisher1996} Fisher, K. B., Bahcall, J. N., Kirhakos, S., \&
Schneider, D. P. 1996, \apj, 468, 469

\reference{foltz1987} Foltz, C. B., Chaffee, F. H. Jr., Hewett, P. C.,
MacAlpine, G. M., et al.\ 1987, \aj, 94, 1423

\reference{foltz1989} Foltz, C. B., Chaffee, F. H. Jr., Hewett, P. C.,
Weymann, R. J., et. al\ 1989, \aj, 98, 1959

\reference{francis1991} Francis, P. J., Hewett, P. C., Foltz, C. B., Chaffee,
F. H., et al.\ 1991, \apj, 373, 465

\reference{francis1992} Francis, P. J., Hewett, P. C., Foltz, C. B., \&
Chaffee, F. H. 1992, \apj, 398, 476

\reference{francis1993a} Francis, P. J. 1993a, \apj, 405, 119

\reference{francis1993b} Francis, P. J. 1993b, \apj, 407, 519

\reference{francis1996} Francis, P. J. 1996, Publ. Astron. Soc. Aust., 13, 212

\reference{french1983} French, H. B., \& Gunn, J. E. 1983, \apj, 269, 29

\reference{jackson1998} Jackson, N., Helbig, P., Browne, I., Fassnacht,
C. D., et al.\ 1998, \aap, 334, L33

\reference{hagen1996} Hagen, H.-J., Hopp, U., Engels, D., \& Reimers, D.
1996, \aap, 308, L25

\reference{hawkins1997} Hawkins, M. R. S., Clements, D., Fried, J. W.,
Heavens, A. F., et al.\ 1997, \mnras, 291, 811

\reference{hewett1989} Hewett, P. C., Webster, R. L., Harding, M. E.,
Jedrzejewski, R. I., et al.\ 1989, \apjl, 346, L61

\reference{hewett1991} Hewett, P. C., Foltz, C. B., Chaffee, F. H., Francis,
P. J., et al.\ 1991, \aj, 101, 1121

\reference{hewett1995} Hewett, P. C., Foltz, C. B., \& Chaffee, F. H. 1995,
\aj, 109, 1498

\reference{hewett1998} Hewett, P. C., Foltz, C. B., Harding, M. E., \& Lewis,
G. F.  1998, \aj, 115, 383.

\reference{hewitt1987} Hewitt, J. N., Turner, E. L., Lawrence, C. R.,
Schneider, D.  P., et al.\ 1987, \apj, 321, 706

\reference{impey1996} Impey, C., Foltz, C. B., Petry, C. E., Browne, I. W. A.,
et al.\ 1996, \apj, 462, L53

\reference{keeton1996} Keeton, C. R., \& Kochanek, C. S. 1996, in {\it
Astrophysical Applications of Gravitational Lensing}, I.A.U.  Symposium No.
173, eds. C. S.  Kochanek \& J. N. Hewitt, p. 419

\reference{keeton1998} Keeton, C. R., Kochanek, C. S., \& Falco, E. E. 1998,
\apj, 509, 561

\reference{kinney1990} Kinney, A. L., Rivolo, A. R., \& Koratkar, A. P. 1990,
\apj, 357, 338

\reference{kochanek1995} Kochanek, C. S. 1995, \apj, 453, 545

\reference{kochanek1996} Kochanek, C. S. 1996, in {\it Astrophysical
Applications of Gravitational Lensing}, I.A.U.  Symposium No. 173, eds. C. S.
Kochanek \& J. N. Hewitt, p. 177

\reference{kochanek1999a} Kochanek, C. S., Falco, E. E., \& Mu\~noz, J. A.,
1999a \apj, 510, 590

\reference{kochanek1999b} Kochanek, C. S., Falco, E. E., Impey, C. D., \& J.
Leh\'ar, McLeod, B. A., et al.\ 1999b, \apj, submitted

\reference{lawrence1984} Lawrence, C. R., Schneider, D. P., Schmidt, M.,
Bennett, C. L., et al.\ 1984, Science, 223, 46

\reference{lehar1999} Leh\'ar, J., Falco, E. E., Impey, C. D., Kochanek, C. S.,
McLeod, B. A., et al.\ 1999, ApJ, in preparation (1999)

\reference{maoz1997} Maoz, D., Rix, H.-W., Gal-Yam, A., \& Gould, A. 1997,
\apj, 486, 75

\reference{mcleod1997} McLeod, B. A. 1997, in 1997 {\it HST Calibration
Workshop}, eds.\ Casertano, S., Jedrzejewski, R., Keyes, T., \& Stevens, M.,
p. 281 

\reference{meylan1989} Meylan, G., \& Djorgovski, S. 1989, \apj, 338, L1

\reference{meylan1990} Meylan, G., Djorgovski, S., Weir, N., \& Shaver, P.,
1990, ESO Messenger, 59, 47

\reference{michalitsianos} Michalitsianos, A. G., Falco, E. E., Mu\~noz, J. A.
\& Kazanas, D., 1997, \apjl, 487, L117

\reference{munoz1998} Mu\~noz, J. A., Falco, E. E., Kochanek, C. S., Leh\'ar,
J., et al.\ 1998, \apjl, 492, L9

\reference{osmer1998} Osmer, P. S. \& Shields, J. C. 1998, astro-ph/9811459

\reference{persson1998} Persson, S. E., Murphy, D. C., Krzeminski, W.,
Roth, M., \& Rieke, M. J. 1998, \aj, 116, 2475

\reference{sanders1989} Sanders, D. B., Phinney, E. S., Neugebauer, G.,
Soifer, B. T., \& Matthews, K. 1989, \apj, 347, 29

\reference{sargent1989} Sargent, W. L. W., Steidel, C. C., \& Boksenberg, A.
1989, \apjs, 69, 703

\reference{schlegel1998} Schlegel, D. J., 
Finkbeiner, D. P. \& Davis, M., 1998, \apj, 500, 525

\reference{schneider1993} Schneider, P. 1993, in {\it Gravitational Lenses in
the Universe}, Proceedings of the 31st Li\`{e}ge International Astrophysical
Colloquium, eds. Surdej, J., Fraipont-Caro, D., Gosset, E., Refsdal, S., \&
Remy, M., p. 41

\reference{small1997} Small, T. A., Sargent, W. L. W., Steidel, C. C. 1997,
\aj, 114, 2254

\reference{smetanka1991} Smetanka, J. J., Bershady, M. A., Kron, R. G., Munn,
J. A., et al.\ 1991, in {\it The Space Distribution of Quasars}, Publications
of the Astronomical Society of the Pacific \#21, ed.\ Crampton, D., 100

\reference{sramek1978} Sramek, R. A. \& Weedman, D. W. 1978, \apj, 221, 468

\reference{steidel1991} Steidel, C. C., \& Sargent, W. L. W. 1991, \aj, 102, 
1610 (SS91)

\reference{steidel1993} Steidel, C. C. 1993, in {\it The Environment and
Evolution of Galaxies}, proceedings of the 3rd Teton Astronomy Conference,
eds. Shull, J. M. \& Thronson, H. A., p. 263

\reference{steidel1995} Steidel, C. C. 1995, in {\it QSO Absorption Lines},
ed.  Meylan, G.\ (Berlin: Springer), p. 139

\reference{steidel1997} Steidel, C. C., Dickinson, M., Meyer, D. M.,
Adelberger, K. L., \& Sembach, K. R. 1997, \apj, 480, 568

\reference{tinney1995} Tinney, C. G. 1995, \mnras, 277, 609

\reference{tinytim} Krist, J., Hook, R. 1997, Tiny Tim Manual, v. 4.4,
Space Telescope Science Institute

\reference{turner1988} Turner, E. L., Hillenbrand, L. A., Schneider, D. P.,
Hewitt, J. N., et al.\ 1988, \aj, 96, 1682

\reference{walsh1979} Walsh, D., Carswell, R. F., \& Weymann, R. J. 1979,
Nature, 279, 381

\reference{wambsganss1995} Wambsganss, J., Cen, R., Ostriker, J. P., \&
Turner, E. L. 1995, Science, 268, 274

\reference{webster1995} Webster, R. L., Francis, P. J., Peterson, B. A.,
Drinkwater, M. J., \& Masci, F. J.\ 1995, Nature, 375, 469

\reference{weedman1982} Weedman, D. W., Weymann, R. J., Green, R. F., \&
Heckman, T. M. 1982, \apjl, 255, L5

\reference{wisotzki1993} Wisotzki, L., K\"{o}ehler, T., Kayser, R., 
\& Reimers, D. 1993, \aap, 278, L15

\reference{wisotzki1995} Wisotzki, L., K\"{o}ehler, T., Ikonomou, M., 
\& Reimers, D. 1995, \aap, 297, L59

\reference{yee1987} Yee, H. K. C., \& Green, R. F. 1987, \apj, 319, 28

\end{references}
\end{document}